\documentstyle[epsf,prl,aps,multicol]{revtex}

 \input{psfig}

\begin{document}
\draft

\title{Interaction of surface acoustic waves with a
two-dimensional electron gas in the presence
of spin splitting of the Landau bands}

\author{I. L. Drichko$^a$, A. M. Diakonov$^a$, V. V.
Preobrazenskiy$^b$, I. Yu. Smirnov$^a$, and A. I. Toropov$^b$ }
\address{$^a$A. F. Ioffe Physico-Technical Institute of Russian
Academy of Sciences, Polytechnicheskaya 26, 194021, St.Petersburg,
Russia;\\ $^b$Semiconductors Physics Institute of Siberian
Division of Russian Academy of Sciences, Ak. Lavrentieva 13,
630090,  Novosibirsk, Russia}

\date{\today}
\maketitle

\begin{abstract}
The absorption and variation of the velocity of a surface acoustic
wave of frequency $f$= 30 MHz interacting with two-dimensional
electrons are investigated in GaAs/AlGaAs heterostructures with an
electron density $n=(1.3 - 2.8) \times 10^{11} cm^{-2}$ at
$T$=1.5 - 4.2 K in magnetic fields up to 7 T. Characteristic features
associated with spin splitting of the Landau level are observed.
The effective g factor and the width of the spin-split Landau
bands are determined: $g^* \simeq 5$ and $A$=0.6 meV. The greater width of
the orbital-split Landau bands (2 meV) relative to the spin-split
bands is attributed to different shielding of the random
fluctuation potential of charged impurities by 2D electrons. The
mechanisms of the nonlinearities manifested in the dependence of
the absorption and the velocity increment of the SAW on the SAW
power in the presence of spin splitting of the Landau levels are
investigated.
\end{abstract}


\begin{multicols}{2}

\section{Introduction}

In a magnetic field $H$ the energy spectrum of a two-dimensional
(2D) electron gas represents a set of discrete levels with
energies

\begin{equation}
  E_N=(N+1/2)\hbar\omega_c \pm (1/2)(g_o\mu_B H+E_{ex}),
  \label{eq1}
\end{equation}

The first term describes the orbital splitting of the Landau
levels $N$ enumerates the Landau levels, $\hbar\omega_c\equiv\hbar
eH/m^*c$ is the cyclotron energy, $g_0\mu_B H$ is the Zeeman
splitting energy, $g_0$ is the $g$ factor in bulk GaAs, and
$\mu_B$ is the Bohr magneton. In the 2D case electron-electron
interactions increase the initial spin splitting {relative lo the
3D case). As a result, (\ref{eq1}) acquires an additional term
$E_{ex}$, which describes the exchange interaction of electrons at
the Landau level. The presence of exchange interaction is
equivalent to an increase in the equivalent $g$ factor \cite{1}
$$g^*\mu_B H=g_0\mu_B H+E_{ex}^0(n\uparrow-n\downarrow).$$ It has been
observed experimentally \cite{2} that the $g$ factor in a
GaAs/AlGaAs heterostructures is $|g^*|$=6.23, which is an order of
magnitude higher than thee value $|g_0|$=0.44 in bulk GaAs. An
increase in the $g$ factor in 2D objects relative to their bulk
counterparts has also been observed in Si/SiO$_2$ \cite{3},
GaInAs/AlInAs \cite{4}, InAs/AlSb/GaSb \cite{5}, and AlAs \cite{6}
heterostructures.

It is evident from (\ref{eq1}) that the energy $E_{ex}$ depends on
the relative population of two spin states, therefore, if the
Fermi level is situated between Landau levels of order ($N,
\uparrow$) and ($N, \downarrow$) the difference
($n\uparrow-n\downarrow$) is a maximum, and the quantity $g^*$
assumes a certain maximum value. But if the Fermi level is between
Landau levels of order ($N, \downarrow$) and ($N+1, \uparrow$),
then $g^*$ assumes a minimum value equal to $g_0$. Consequently,
the value of $g^*$ is an oscillating function of the magnetic
field \cite{1}. Indeed, oscillations of the $g^*$ factor in
magnetic field have been observed in \cite{5}.

In this paper we report measurements of the absorption $\Gamma$
and velocity increments $\Delta V/V$ of surface acoustic waves
(SAWs) in a piezoelectric: material as a result of their
interaction with 2D electrons in GaAs/AlGaAs heterostructures a
magnetic field in the presence of spin splitting of the Landau
levels.

\section{EXPERIMENTAL PROCEDURE}

The details of the procedure used to measure the SAW absorption
are described in \cite{7}. Here we mention only that the SAWs were
generated by means of an interdigital transducer situated on the
surface of a $LiNbO_3$ (lithium niobate) piezoelectric insulator,
into which an rf pulse (30-150 MHz) of length 0.5 $\mu sec$ and
repetition rate 50 Hz was propagated. This pulse was chopped from
the continuous output signal of a microwave oscillator. The signal
transmitted through the sample was received by an analogous
transducer formed on the same surface.

A modified version of the pulse-interference method was used to
measure $\Delta V/V$. The duty phases of two pulses were compared
in a phase detector the SAW-generating pulse on the lithium
niobate surface and the pulse arriving at the receiving
transducer, attenuated by interaction with 2D electrons. The error
signal at the output of the phase detector was sent to the
SAW-generating oscillator and thus altered its frequency in such a
way as to reduce the difference between the duty phases of the
indicated pulses to zero. This tracking system was operative
throughout the entire experiment with a frequency meter
continuously tracking the phase difference between the indicated
pulses. The frequency change recorded by this technique was
converted into the corresponding velocity change.

The sample was held firmly against the lithium niobate surface by
a spring, where the distance of the 2D channel from the lithium
niobate surface was shorter than the SAW wavelength ($\lambda
=100\mu m$ at $f=30MHz$). With this mounting of the sample a random
gap $a$ is formed between it and the lithium niobate surface, its
width ($\sim 0.5 \mu m$) most likely being governed by the
roughnesses of the surfaces of the sample and the lithium niobate.
The SAW-induced strain in the sample is not transferred in our
experimental configuration. The alternating electric field
accompanying the SAW penetrates into the channel carrying 2D
electrons, inducing currents in the 2D channel and, accordingly.
Joule losses, causing not only SAW energy to be absorbed, but also
changing the velocity of the wave. The receiving transducer
detects the amplitude of the SAW signal transmitted through the
sample. This procedure can therefore be used to introduce an
alternating electric field into the sample without having to use
contacts. The absorption $\Gamma$ and velocity increment $\Delta
V/V$ were measured in a vacuum chamber in a magnetic field up to 7
T at temperatures of 1.5-4.2 K in the linear regime (the acoustic
power did not exceed $10^{-6} W$) and at T=1.5K in the
measurements, depending on the acoustic power.

The GaAs/AlGaAs heterostructures were fabricated by molecular-beam
epitaxy with carrier densities $n=(1.3-2.8)\times 10^{11}cm^{-2}$
and mobilities $\mu=(1-2)\times 10^5 cm^2/(V \cdot s)$. The
density and mobility of 2D electrons were determined by a
contactless acoustical method \cite{8}.

\section{EXPERIMENTAL RESULTS AND DISCUSSION}
\subsection{Linear regime}

In our experimental configuration the absorption coefficient
$\Gamma$ and the velocity increment $\Delta V/V$ are given by the
equations \cite{9}

\begin{eqnarray}
\Gamma=8.68 \frac{K^2}{2} qA\frac
{ (\frac{4\pi \sigma_{1}} {\varepsilon_sV})t(q) } 
{[1+(\frac{4\pi \sigma_{2}}{\varepsilon_s V})t(q)]^2+
[(\frac{4\pi \sigma_{1}}{\varepsilon_s V})t(q)]^2}, 
\frac{dB}{cm}
\label{eq2}
\end{eqnarray}
$$A=8b(q)(\varepsilon_1+\varepsilon_0)\varepsilon_0^2\varepsilon_s
e^{(-2q(a+d))},$$ $$ \frac{\Delta V}{V}= \frac{K^2}{2}
A\frac{(\frac{4\pi \sigma_{2}}{\varepsilon_s V})t(q)+1}
{[1+(\frac{4\pi \sigma_{2}}{\varepsilon_s V})t(q)]^2+ [(\frac{4\pi
\sigma_{1}}{\varepsilon_s V})t(q)]^2}, $$
$$b(q)=[b_1(q)[b_2(q)-b_3(q)]]^{-1,}$$
$$t(q)=[b_2(q)-b_3(q)]/[2b_1(q)],$$
$$b_1(q)=(\varepsilon_1+\varepsilon_0)(\varepsilon_s+\varepsilon_0)
-(\varepsilon_1-\varepsilon_0)
(\varepsilon_s-\varepsilon_0)e^{(-2qa)},$$
$$b_2(q)=(\varepsilon_1+\varepsilon_0)(\varepsilon_s+\varepsilon_0)+
(\varepsilon_1+\varepsilon_0)(\varepsilon_s-\varepsilon_0)e^{(-2qd)},$$
\begin{eqnarray}
b_3(q)=(\varepsilon_1-\varepsilon_0)(\varepsilon_s-\varepsilon_0)e^{(-2qa)}
+(\varepsilon_1-\varepsilon_0)(\varepsilon_s+\varepsilon_0)\times
\nonumber & \\ \times e^{[-2q(a+d)]}, \nonumber
\end{eqnarray}
where $K^2$ is the electromechanical coupling constant of
$LiNbO_3$, $q$ and $V$ are the wave vector and velocity of the
SAW, respectively, $a$ is the distance between the insulator and
the investigated heterostructure, $d$ is the depth of the 2D
layer, $\varepsilon_1$, $\varepsilon_0$, and $\varepsilon_s$, are
the dielectric constants of lithium niobate, vacuum, and gallium
arsenide, respectively, and $\sigma_1$ and $\sigma_2$ are the
components of the rf conductivity of 2D electrons, which is
complex-valued: $\sigma_{xx}^{hf}=\sigma_1-i\sigma_2$ \cite{10}.
These equations can be used to determine the values of $\sigma_1$
and $\sigma_2$ from the experimentally measured quantities
$\Gamma$ and $\Delta V/V$.

It is evident from (\ref{eq2}) that the dependence of $\Gamma$ and
$\Delta V/V$ on the magnetic field $H$ is determined by the $H$
dependence of the components of the dissipative conductivity
$\sigma_{xx}$, so that the quantization of the electron spectrum
in a magnetic field, inducing Shubnikov-de Haas resistance
oscillations, also generates oscillations in our measured effects.
Figure 1 shows experimental curves of $\Gamma$ and $\Delta V/V$ as
functions of the magnetic field for samples AG49 ($n=2.7 \times
10^{11} cm^{-2}$), AG106 ($n=1.3 \times 10^{11} cm^{-2}$), and
BP92 ($n=2.8 \times 10^{11} cm^{-2}$) at $T$=4.2K and 1.5K. It is
evident from the figures that $\Gamma$ and $\Delta V/V$ oscillate
in a magnetic field; additional peaks, nonexistent or faint at
$T$=4.2K, are observed at $T$=1.5K for $H$=2.2T and 3.65 T (AG49),
$H$=2.4T and 4T (BP92), and $H$=5.5T (AG106).

The emergence of these peaks is associated with spin splitting of
the Landau zones, because for samples AG49 and BP92 their magnetic
field positions correspond to occupation numbers $\nu=3$ and
$\nu=5$, whereas for sample AG106 they correspond to $\nu=1$,
where
$$\nu=nch/eH,$$

$c$ is the speed of light, $h$ is Planck's constant, and $e$ is
the electron charge. We are convinced that the SAW absorption peak
in a magnetic field corresponding to spin splitting of the Landau
level has been observed in \cite{11}, but was not identified by
the authors.

Regarding the temperature dependence, in sample AG49 the
absorption peaks corresponding to $\nu=3$ are not observed at $T$=
4.2 K and begin to appear at $T$ = 3.2 - 3.7 K for different
mountings of the sample, while the peak corresponding to $\nu=5$
is observed only at $T$=1.5K. The energy of spin splitting of the
Landau levels is $E_g=g^*\mu_B H$, so that phenomena associated
with spin splitting can occur only when the condition $E_g > kT$
holds; consequently, the smaller the value of $\nu$, the higher is
the temperature at which they are observed.

The profile of the curves for $\Gamma (H)$ in strong magnetic
fields (splitting in two of the $\Gamma$ peak in the vicinity of
even occupation numbers) is attributable to the relaxation
behavior of the absorption and is analyzed in detail in \cite{7}.
Moreover, it is evident from Fig.1 that at $T$=1.5 K the value of
the $\Gamma$ peak for odd occupation numbers, $\nu=1$ (AG106) and
$\nu=3$ (AG49) is higher (in spin splitting) than the maximum
value of the absorption peak for even values $\nu=2, 4,6$, which
correspond to orbital splitting. This fact is very important in
regard to understanding the nature of the interaction of SAWs with
2D electrons in heterostructures in the regime of the quantum Hall
effect. And indeed it has been reported in several papers (e.g.,
Refs. 11 and 12) that the experimentally measured value of
$\Gamma$ for all magnetic fields is given by an equation of the
type (\ref{eq2}), where the role of $\sigma$ can be taken by the
conductivity $\sigma_{xx}^{dc}$ measured for a direct current.
Here the maximum absorption $\Gamma^{max}$ does not depend on
$\sigma_{xx}$, i.e., is universal. It has been shown \cite{7} that
the dependence $\Gamma(H)$ is described by the conductivity
determined from static measurements only when the 2D electrons are
delocalized. In the regime of the integer-valued quantum Hall
effect the Fermi level is situated halfway between two consecutive
Landau levels, the electrons are localized, and the dc and ac
conductivity mechanisms differ, so that $\sigma_{xx}^{rf} >
\sigma_{xx}^{dc}=0$. In this case allowance must be made for the
fact that $\sigma_{xx}^{rf}$ has a complex form. It is difficult
to analyze the conditions for attaining the maximum $\Gamma^{max}$
according to (\ref{eq2}) as a function of the magnetic field in
this case, but experiment shows that it is achieved when $Re
\sigma_{xx}=\sigma_{1}\approx Im \sigma_{xx}=\sigma_2$. The
maximum $\Gamma^{max}$ can then be calculated according to
(\ref{eq2}) with $\sigma_1=\sigma_2=\sigma$, and if, as experiment
shows, $4\pi\sigma/\varepsilon_s V > 1$ in this case, the maximum
absorption is $\Gamma^{max}\simeq 1/2\sigma$, i.e., depends on the
conductivity.

To understand the nature of the magnetic field dependence of
$\Delta V/V$, we must look once again at the physical phenomena
associated with interaction between SAWs and 2D electrons. For
$H=0$ electrons are delocalized in the investigated samples, and
the 2D electron gas has a high conductivity, so that the rf
electric fields accompanying SAW propagation are completely
shielded. Here the SAW velocity in lithium niobate has a certain
value $V$. In magnetic fields corresponding to the position of the
Fermi level between two consecutive Landau levels the conductivity
in the 2D channel diminishes, and the sample essentially becomes
an insulator; the shielding of the piezoelectric fields in lithium
niobate decrease in this event, disappearing altogether at the
conductivity minimum. According to theory, \cite{13,14}
piezoelectric materials have a sound velocity increment, which
increases the velocity by an amount proportional to the
electromechanical coupling constant $K^2$, so that for a
piezoelectric material the SAW velocity $V_0 > V$. The quantity
$\Delta V/V=(V_0 - V)/ V$ is measured in experiment. It is evident
from Fig.1 that for all the samples the inequality $(\Delta
V/V)_{spin} < (\Delta V/V)_{orb}$ always holds in the entire
temperature range. This means that the conductivity of the 2D
electron system in spin splitting is always higher than in orbital
splitting. If we compare the $(\Delta V/V)_{spin}$ curves for
samples AG49 and BP92, which have close electron densities, we
find that for $\nu=3$ the second sample does not exhibit the
$(\Delta V/V)_{spin}$ peak. This means that the conductivity of
the 2D channel is higher in the second sample than in the first.

To analyze the results, it is more practical to work with the rf
conductivity $\sigma_{s}$ in magnetic fields corresponding to
occupation numbers $\nu$= 1 and 3, which is calculated from the
experimentally measured quantities $\Gamma$ and $\Delta V/V$. For
this purpose it is necessary to know the random gap a between the
sample and the surface of the piezoelectric substrate. The gap $a$
can be determined from the solution of (\ref{eq3}). If we assume
that in relatively weak magnetic fields, where electrons are
delocalized \cite{8}, the quantity
$\sigma_{xx}=\sigma_1=\sigma_{xx}^{dc}$ does not depend on the
frequency, while $\sigma_2=0$ and $4\pi\sigma_{xx}/\varepsilon_sV
\gg 1$, we then have $\Gamma\sim1/\sigma_{xx}$; we now infer from
(\ref{eq2}) that the ratio of the absorption coefficients in these
fields at two different frequencies for the same sample mounting
(i.e., for the same gap) can be written in the form

\begin{equation}
  \frac{\Gamma(q_1)}{\Gamma(q_2)}=
  \frac{[q_1b(q_1)t(q_2)]}{[q_2 b(q_2)t(q_1)]}e^{-2(a+d)(q_1-q_2)}.
  \label{eq3}
\end{equation}

Here $q_1$ and $q_2$ are wave vectors corresponding to two
different SAW frequencies. In our experiments we have found that
$a \simeq 0.5 \mu m$ for different sample mountings.

Consequently, from acoustical measurements we have calculated $Re
\sigma_{sp}=\sigma_1$ and $Im \sigma_{sp}=\sigma_2$, along with their
dependences on the temperature, the magnetic field, and the SAW
power. At $T$=1.5 K we find that $\sigma_1/\sigma_2$ has the values
9.7 ($\nu$=3) for sample AG49, 17 ($\nu=3$) for BP92, and 0.9
($\nu=1$) for AG106.

As mentioned, the maximum value of the absorption coefficient
$\Gamma^{max}$ is attained for $\sigma_1=\sigma_2$. Consequently,
if $\sigma_1 > \sigma_2$ in a magnetic field corresponding to the
absorption peak, the maximum absorption $\Gamma^{max}$ is still
not attained. This case occurs in magnetic fields with $\nu>8$ in
all the samples, and in sample AG49 it also occurs for $\nu=3$,
even at $T$=1.5K. Experiment shows that the conductivity
$\sigma_1=\sigma_2=\sigma$ at which the absorption attains its
maximum value $\Gamma^{max}$ is greater for even values of $\nu$
than for odd values, which correspond to spin splitting, and since
$\Gamma^{max}\sim1/\sigma$, the maximum absorption for spin
splitting is found to be greater than for orbital splitting. This
result is clearly evident in Fig.1 for sample AG106.

The temperature dependence of $Re \sigma_{sp}\equiv \sigma_1 (\nu=
1;3)$ for all the samples in the investigated temperature range is
well described by the law

\begin{equation}
  Re \sigma_{sp}= \sigma_1 \sim exp(-E_g/2kT).
  \label{eq4}
\end{equation}

This law is confirmed by the linearity of the plots of $ln
\sigma_1$ as a function of 1/$T$ ($f$=30MHz) shown in Fig.2 for all
the investigated samples. From the slopes of these lines we have
determined the activation energies $E_g=g^*\mu_BÍ$, which are
determined by the spin splitting energy. The inset in Fig.2 shows
the magnetic field dependence of $E_g$. It is evident that $E_g$
is a linear function of the magnetic field, so that the g* factor
can be determined from the slope of $E_g(H)$, $g^*$=5. This value
agrees with the results of other studies \cite{2,15,16}. It is
evident from the figure that the $E_g (H)$ line does not pass
through the origin when extrapolated to $H$=0, probably because
the Landau level broadens as a result of the impurity fluctuation
potential \cite{15}. We have determined the width of the
spin-split Landau levels from the intercept of the $E_g(H)$ line
with the energy axis at $H$=0: $A$=0.58meV. We have previously
\cite{17} determined the widths of the bands in the case of
orbital splitting for the same samples: $A\approx $2meV (AG49).
Consequently, the width of the Landau bands is greater in orbital
splitting than in spin splitting. As mentioned above, the
conductivity of the 2D electron system is always greater in spin
splitting than in orbital splitting (for low occupation numbers).
Accordingly, the greater the conductivity, the more effective is
the shielding of the impurity fluctuation potential and, as a
result, the smaller is the width of the band.

\subsection{Nonlinear regime}

Figure 3 shows the dependence of $\Gamma_{spin}$ on the rf source
output power $P$ ($f$=30MHz) for samples AG49 ($\nu$=3) and AG106
($\nu$=l) at $T$=1.5 K. It is evident from the figure that as the
power is increased, the absorption associated with spin splitting
of the Landau band, $\Gamma_{spin}$, decreases and becomes equal
to zero at a certain power level. The relatively small value of
$\Delta V/V$ for sample AG49 for $\nu$=3, in contrast with
$\nu$=2;4, implies that the conductivity of the 2D system is
already fairly high, i.e., a large number of delocalized electrons
is present. In this case the behavior of $\Gamma_{spin}(P)$ can be
attributed to heating of the electron gas by the SAW electric
field, where $\Gamma_{spin}(P) \rightarrow 0$ as
$kT_e \rightarrow g\mu_B H$. To describe the heating of the electron
gas, we need to know the electron temperature $T_e >T$ ($T$ is the
temperature of the lattice), which can be determined by comparing
the $\Gamma(P)$ and $\Gamma(T)$ curves. The SAW electric field
that penetrates into the channel containing the 2D electron gas is
given by the expression

\begin{eqnarray}
\  |E|^2=K^2\frac{32\pi}{V}(\varepsilon_1+\varepsilon_0)
\frac{zqe^{(-2q(a+d))}} {(1+\frac{4\pi \sigma_{2}}{\varepsilon_s
V}t)^2+(\frac{4\pi \sigma_{1}}{\varepsilon_s V}t)^2}W,\label{eq5}
\end{eqnarray}

\begin{eqnarray}
z=[(\varepsilon_1 + \varepsilon_0)(\varepsilon_s + \varepsilon_0)-
e^{(-2qa)}(\varepsilon_1-\varepsilon_0) \nonumber & \\ \times
(\varepsilon_s-\varepsilon_0)]^{-2}, \label{eq6}
\end{eqnarray}

where $W$ is the input SAW power normalized to the width of the
sound track. The energy losses in this case are
$Q=e \mu E^2=4W \Gamma$ \cite{18}. The experimental plot of $Q(T_e)$
is shown in Fig. 4. It is found to be well described by the
function $Q=A_5(T^5_e - T^5)$. A similar dependence has been
obtained in an investigation of nonlinearities in weak magnetic
fields, when the electrons exist in delocalized states \cite{18},
corresponding to the relaxation of energy at piezoacoustic phonons
in the presence of strong shielding \cite{19}. The value obtained
for the coefficient $A_5$ from this experiment is 25$eV/(s\cdot
K^5)$, in contrast with the theoretical value deter-mined from
equations in \cite{19} for this sample: 62 $eV/(s\cdot K^5)$. The
difference in the coefficients can be attributed to errors in the
determination of the absolute value of the SAW power.

In sample AG106 ($\nu$=1) the same mechanism could not account for
the behavior of $\Gamma_{spin}(P)$. Indeed, we can infer from the
profile of the peak of the absorption coefficient and the
increment $\Delta V/V$ that the conductivity in this sample at
$T$= 1.5 K is smaller than in sample AG49, where the splitting in
two of the absorption peak indicates the localization of carriers
situated in the upper band with an oppositely directed spin
(relative to the lower band). In this situation we can assume that
the nonlinear effects are associated with a decrease in the
activation energy in the SAW electric field \cite{20} for
electrons existing in localized states at the Fermi level
(Poole-Frenkel effect). Now the dependence of the real part of the
conductivity on the SAW electric field is given by the expression

\begin{equation}
  \sigma_1=Re\sigma_{spin} \propto n(E)=n_0exp(2e^{3/2}E^{1/2}
  \varepsilon_s^{-1/2}/kT),
  \label{eq7}
\end{equation}
where $n_0$ is the carrier density in the upper Landau band in the
linear regime at $T$=1.5 K. The linear behavior of $ln\sigma_1$ as a
function of $E^{1/2}$, with slope 10 $(cm \cdot s^2/g)^{1/4}$,
corroborates this assumption (Fig.5). The slope calculated from
(\ref{eq7}) is 28 $(cm \cdot s^2/g)^{1/4}$.

We assume that both of the above-mentioned effects responsible for
the dependencies $\Gamma (P)$ and $\Delta V/V(P)$ actually
coexist, but when delocalized electrons dominate the upper,
spin-split-off band, heating of the 2D electron gas plays a
greater role; on the other hand, if free electrons are few in
number, the dominant mechanism of nonlinearity at first is a
reduction of the activation energy in the SAW electric field,
causing the number delocalized electrons to increase in the upper
band of the spin-split Landau bands, which are heated by the SAW
electric field.

\section{CONCLUSIONS}

In our investigations of the absorption and velocity increment of
surface acoustic waves ($f$=30MHz) due to interaction with
two-dimensional electrons in GaAs/AlGaAs heterostructures (with
electron densities $n = 1.3\times10^{11} cm^{-2}$, $n =
2.7\times10^{11} cm^{-2}$, and $n = 2.8\times10^{11} cm^{-2}$ at
$T$=1.5-4.2 K in magnetic fields up to 7T, we have:

\begin{itemize}

\item observed peaks associated with spin splitting of the Landau
levels;

\item evaluated the effective g factor, $g^*$=5.

\item determined the width of the Landau bands associated with spin
splitting: 0.6meV, which is found to be smaller than the width of
the Landau bands for orbital splitting: 2 meV; we have shown that
the conductivity of the 2D electron system in spin splitting is
always greater than in orbital splitting, so that the fluctuation
potential of charged impurities, which governs the width of the
Landau bands, is shielded more effectively in spin splitting;

\item investigated the mechanisms of the nonlinearities manifested
in the dependencies of the absorption coefficient and the SAW
velocity increment on the SAW power in the presence of spin
splitting of the Landau levels.

\end{itemize}

\section{Acknowledgements}

The authors are grateful to V. D. Kagan for many discussions, to
A. V. Suslov for helping with the measurements, and to D. A.
Pristinski for carrying out the numerical computations.

This work has been supported by grants from the Russian Fund for
Fundamental Research, RFFI Grant No. 98-02-18280, and from the
Ministry of Science, No. 97-1043.

}
\end{multicols}

\newpage

\begin{figure}[t]
\centerline{\psfig{figure=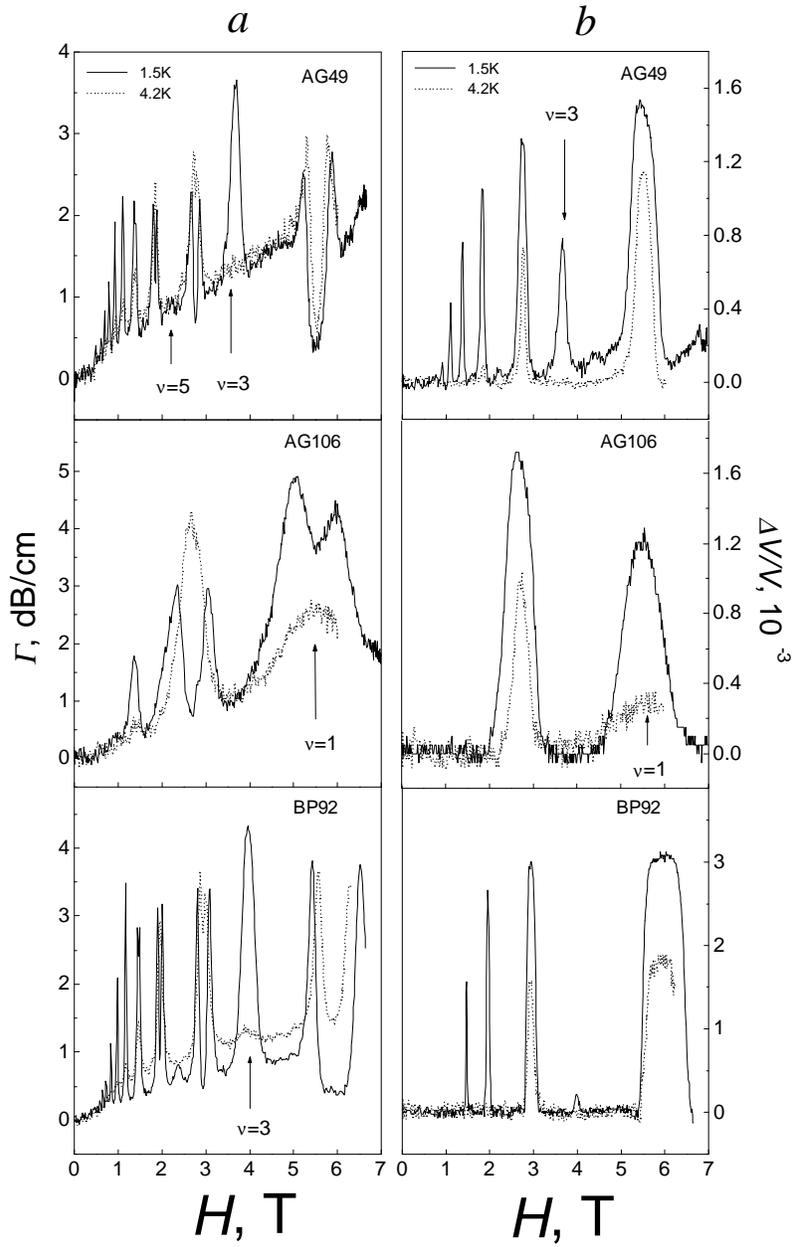}} 
\caption{ Dependence of the SAW absorption coefficient $\Gamma$
(a) and the relative SAW velocity increment $\Delta V/V$ (b) on
the magnetic field $H$ for different samples at temperatures of
4.2 K and 1.5 K; the SAW frequency is $f$=30MHz. \label{fig1}}
\end{figure}

\newpage

\begin{figure}[h]
\centerline{\psfig{figure=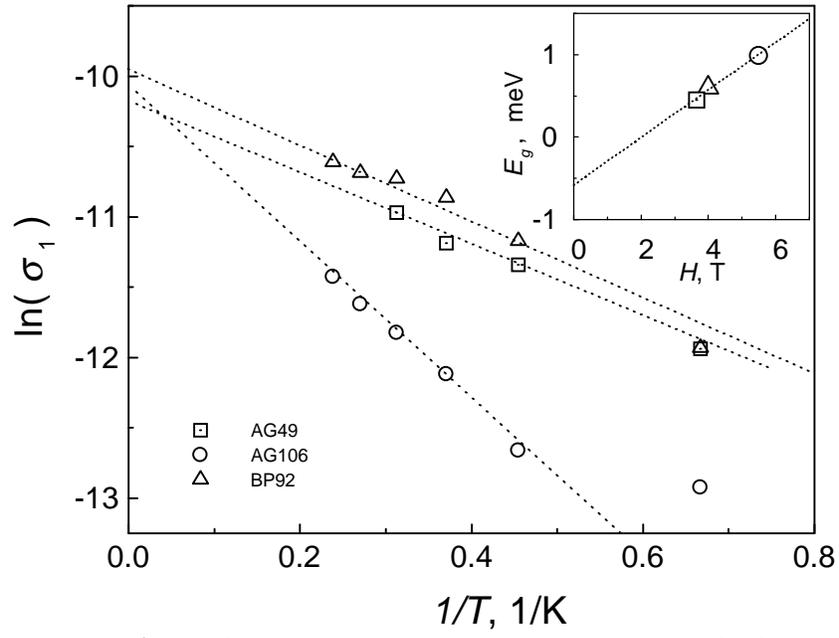}} 
\caption{Dependence of $ln\sigma_1$ on 1/$T$ for 
different samples. Inset: dependence of the activation
energy $E_g$, on the magnetic field. \label{fig2}}
\end{figure}

\newpage

\begin{figure}[h]
\centerline{\psfig{figure=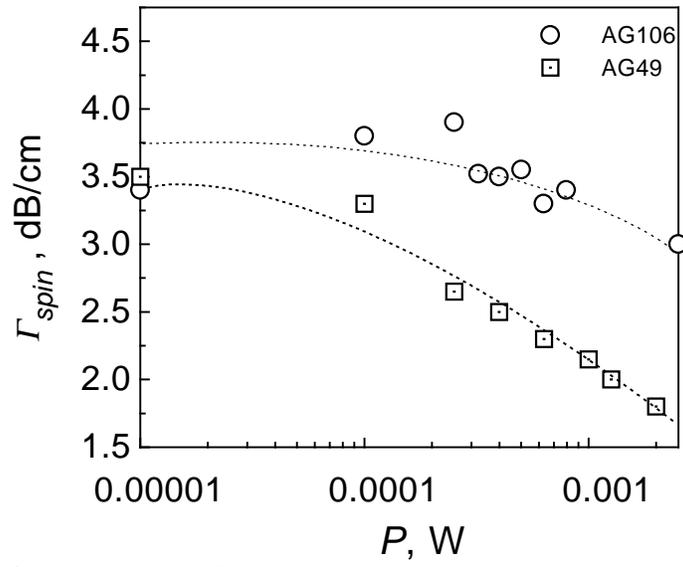}} 
\caption{Dependence of the SAW absorption coefficient
$\Gamma_{spin}$ on the generator output power $P$ for samples AG49
and AG106 at $T=$1.5 K. \label{fig3}}
\end{figure}

\newpage

\begin{figure}[h]
\centerline{\psfig{figure=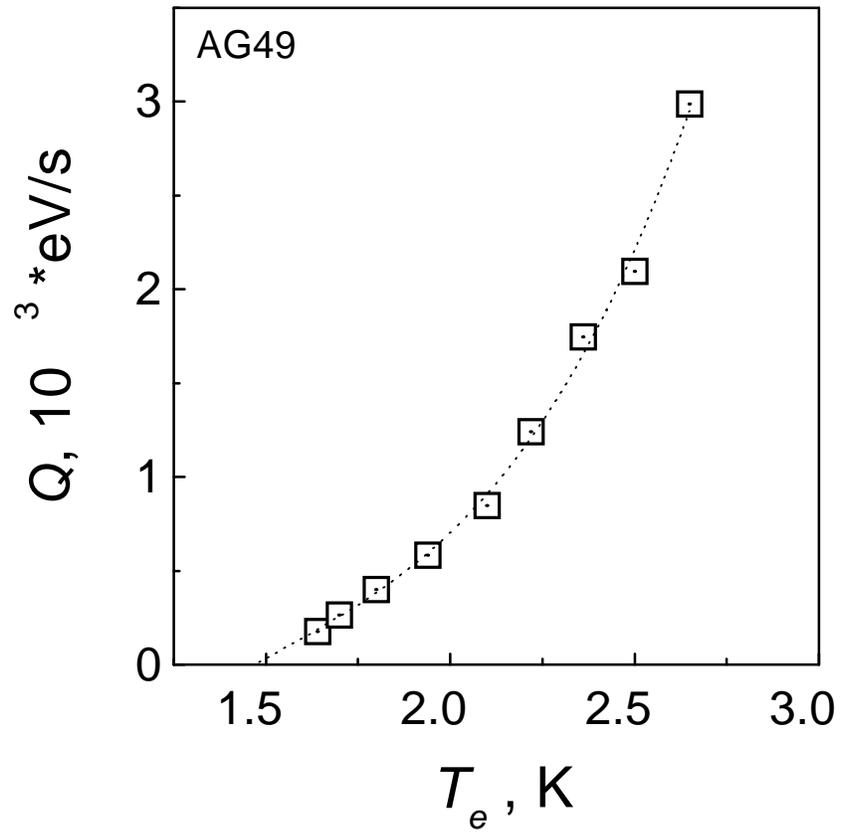}} 
\caption{Dependence of the energy losses $Q$ on the electron
temperature $T_e$ for sample AG49 at $T=$1.5 K. \label{fig4}}
\end{figure}

\newpage

\begin{figure}[h]
\centerline{\psfig{figure=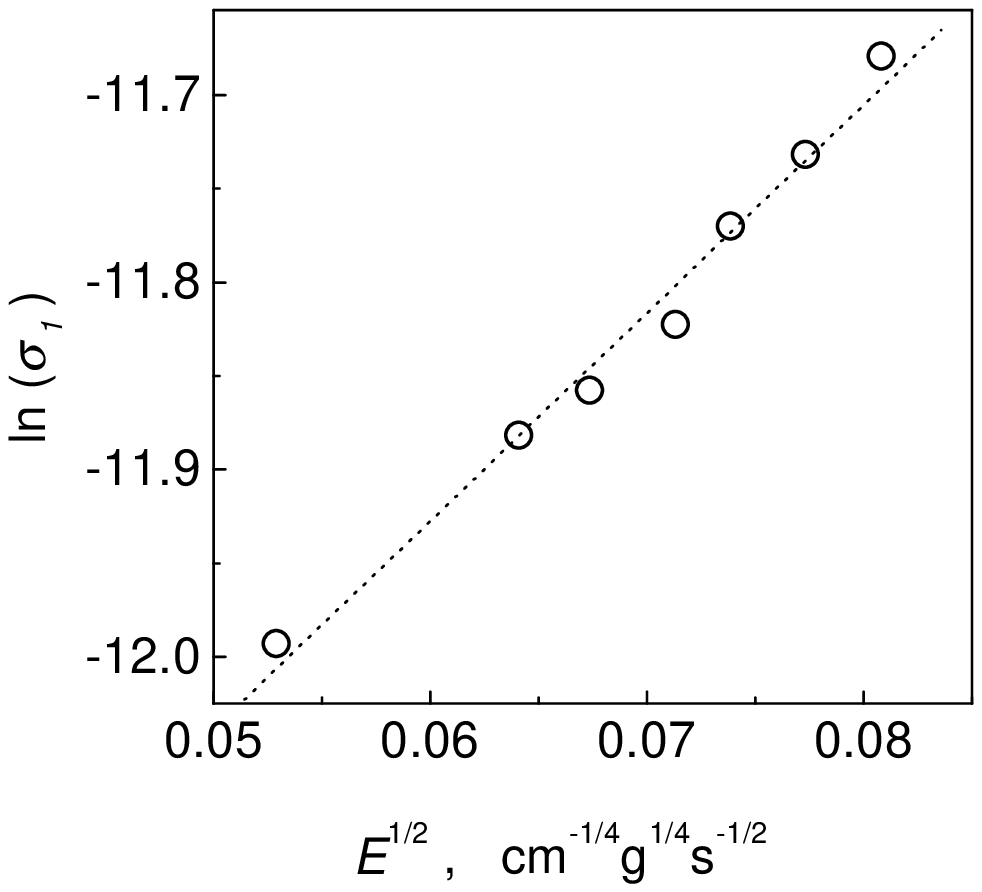}} 
\caption{Dependence of $ln\sigma_1$, on $E^{1/2}$ for sample AG106
at $T=$1.5 K.\label{fig5}}
\end{figure}


\begin{references}

\bibitem{1}T. Ando and Y. Uemura, J. Phys. Soc. Jpn. 37, 1044 (1974).
\bibitem{2}R. J.Nicholas, R. J. Haug, K. V. Klitzing, and G. Weimann, Phys. Rev. Â
37.1294 (1988).
\bibitem{3}F. F. Fang and P. J. Stiles, Phys. Rev. 174, 823
(1968).
\bibitem{4}R. J. Nicholas, M. A. Brummel, J. C. Portal, K. Y. Cheng,
A. Y. Cho, and T. P. Pearsall, Solid State Common. 45. 911 (1983).
\bibitem{5}E. E. Mendez, J. Nocera, and W. 1. Wang, Phys. Rev. B 47, 13 937
(1993).
\bibitem{6}S. P. Papadakis, E. P. de Poorte, and M. Shayegan,
cond-mat/9808158.
\bibitem{7}I. L. Drichko,  A. M. Diakonov,
A. M. Kreshchuk, T. A. Polyanskaya, I. G. Savel'ev, I. Yu.
Smirnov, and A. V. Suslov, Fiz. Tekh. Poluprovodn 31, 451 (1997)
[Semiconductors 31. 384 (1997)].
\bibitem{8}I. L. Drichko and I. Yu. Smirnov,
Fiz. Tekh. Poluprovodn 31, 1092 (1997) [Semiconductors 31,933
(1997)].
\bibitem{9}V. D. Kagan, Fiz. Tekh. Poluprovodn. 31. 470 (1997)
[Semiconductors 31,407 (1997)].
\bibitem{10}I. L. Aleiner and B. I.
Shklovskii, Int. J. Mod. Phys. B 8, 801 (1994).
\bibitem{11}F. Guillion, A. Sachrajda, M. D'Iorio, R. Boulet, P.
Coleridge, Can. J. Phys.
69,461 (1992).
\bibitem{12}A. Wixforth, J. Scriba, M. Wassermeier, J. P.
Kotthaus, G. Weimann and W. Schlapp, Phys. Rev. B 40. 7874
(1989).
\bibitem{13}V. L. Gurevich, Fiz. Tverd. Tela (Leningrad) 4. 909
(1962) [Sov. Phys. Solid State 4. 668 (1962)].
\bibitem{14}A. R. Hutson and
D. L. White, J. Appl. Phys. 33. 40 (1962).
\bibitem{15}A. Usher, R. J.
Nicholas, J. J. Hams, and C. T. Foxon, Phys. Rev. B 41,
1129(1990).
\bibitem{16}D. R. Leadley, R. J. Nicholas, J. J. Hams, and C. T.
Foxon, cond-mat/9805332.
\bibitem{17}I. L. Drichko, A. M.
Diakonov, V. D. Kagan, I. Yu. Smirnov, and A. I. Toropov,
Proc. of the 24th ICPS (Jerusalem, Israel)
on CD-ROM, World Publishing, Singapore (1998).
\bibitem{18}I. L. Drichko. A. M. D'yakonov. V. D. Kagan. A. M.
Kreshchuk, T. A. Polyanskaya, I. G. Savel'ev, I. Yu. Smirnov, and
A. V. Suslov. Fiz. Tekh. Poluprovodn. 31, 1357 (1997)
[Semiconductors 31, 1170 (1997)].
\bibitem{19}V. Karpus. Fiz. Tekh.
Poluprovodn. 22,439 (1988) [Sov. Phys. Semicond. 22, 268 (1988)].
\bibitem{20}L. S. Stil'bans. Physics of Semiconductors (in Russian), Sov.
Radio, Moscow (1967).

\end{references}
\end{document}